\documentclass[aps,prl,twocolumn,superscriptaddress,showpacs,preprintnumbers,amsmath,amssymb]{revtex4}

\usepackage{graphicx} 
\usepackage{dcolumn}  

\graphicspath{{ps}}

\begin{document}

\preprint{\vbox{
}}

\title{ \quad\\[0.5cm]  \boldmath Search for $B^0 \to p\overline{p}$, $\Lambda \overline{\Lambda}$ and $B^+\to p \overline{\Lambda}$ at Belle}

\affiliation{Budker Institute of Nuclear Physics, Novosibirsk}
\affiliation{Chiba University, Chiba}
\affiliation{Chonnam National University, Kwangju}
\affiliation{University of Cincinnati, Cincinnati, Ohio 45221}
\affiliation{Gyeongsang National University, Chinju}
\affiliation{University of Hawaii, Honolulu, Hawaii 96822}
\affiliation{High Energy Accelerator Research Organization (KEK), Tsukuba}
\affiliation{Hiroshima Institute of Technology, Hiroshima}
\affiliation{Institute of High Energy Physics, Chinese Academy of Sciences, Beijing}
\affiliation{Institute of High Energy Physics, Vienna}
\affiliation{Institute for Theoretical and Experimental Physics, Moscow}
\affiliation{J. Stefan Institute, Ljubljana}
\affiliation{Kanagawa University, Yokohama}
\affiliation{Korea University, Seoul}
\affiliation{Kyungpook National University, Taegu}
\affiliation{Swiss Federal Institute of Technology of Lausanne, EPFL, Lausanne}
\affiliation{University of Ljubljana, Ljubljana}
\affiliation{University of Maribor, Maribor}
\affiliation{University of Melbourne, Victoria}
\affiliation{Nagoya University, Nagoya}
\affiliation{Nara Women's University, Nara}
\affiliation{National Central University, Chung-li}
\affiliation{National United University, Miao Li}
\affiliation{Department of Physics, National Taiwan University, Taipei}
\affiliation{H. Niewodniczanski Institute of Nuclear Physics, Krakow}
\affiliation{Nihon Dental College, Niigata}
\affiliation{Niigata University, Niigata}
\affiliation{Osaka City University, Osaka}
\affiliation{Osaka University, Osaka}
\affiliation{Panjab University, Chandigarh}
\affiliation{Peking University, Beijing}
\affiliation{Princeton University, Princeton, New Jersey 08544}
\affiliation{University of Science and Technology of China, Hefei}
\affiliation{Seoul National University, Seoul}
\affiliation{Sungkyunkwan University, Suwon}
\affiliation{University of Sydney, Sydney NSW}
\affiliation{Tata Institute of Fundamental Research, Bombay}
\affiliation{Toho University, Funabashi}
\affiliation{Tohoku Gakuin University, Tagajo}
\affiliation{Tohoku University, Sendai}
\affiliation{Department of Physics, University of Tokyo, Tokyo}
\affiliation{Tokyo Institute of Technology, Tokyo}
\affiliation{Tokyo Metropolitan University, Tokyo}
\affiliation{Tokyo University of Agriculture and Technology, Tokyo}
\affiliation{University of Tsukuba, Tsukuba}
\affiliation{Virginia Polytechnic Institute and State University, Blacksburg, Virginia 24061}
\affiliation{Yonsei University, Seoul}
 \author{M.-C.~Chang}\affiliation{Tohoku University, Sendai}\affiliation{Department of Physics, National Taiwan University, Taipei} 
  \author{K.~Abe}\affiliation{High Energy Accelerator Research Organization (KEK), Tsukuba} 
  \author{K.~Abe}\affiliation{Tohoku Gakuin University, Tagajo} 
  \author{I.~Adachi}\affiliation{High Energy Accelerator Research Organization (KEK), Tsukuba} 
  \author{H.~Aihara}\affiliation{Department of Physics, University of Tokyo, Tokyo} 
  \author{Y.~Asano}\affiliation{University of Tsukuba, Tsukuba} 
  \author{T.~Aushev}\affiliation{Institute for Theoretical and Experimental Physics, Moscow} 
  \author{S.~Bahinipati}\affiliation{University of Cincinnati, Cincinnati, Ohio 45221} 
  \author{A.~M.~Bakich}\affiliation{University of Sydney, Sydney NSW} 
  \author{I.~Bedny}\affiliation{Budker Institute of Nuclear Physics, Novosibirsk} 
% ...
\collaboration{The Belle Collaboration}

\begin{abstract}
We report results of a search for the charmless two-body baryonic decays $B^0 \to p \overline{p}$, $B^0 \to \Lambda \overline{\Lambda}$, and $B^+\to p \overline{\Lambda}$ based on the analysis of a $140\,{\rm fb}^{-1}$ data sample. We set 90\% confidence level upper limits on their branching fractions: $\mathcal {B}$($B^0 \to p \overline{p}$) $<$ 4.1 $\times 10^{-7}$, 
$\mathcal {B}$($B^0 \to \Lambda \overline{\Lambda}$) $<$ 6.9 $\times 10^{-7}$, and
$\mathcal {B}$($B^+ \to p \overline {\Lambda}$) $<$ 4.9 $\times 10^{-7}$.
\end{abstract}

\maketitle

\tighten

{\renewcommand{\thefootnote}{\fnsymbol{footnote}}}
\setcounter{footnote}{0}

Motivated by the recent observation of the two-body baryonic decay $\overline{B}{}^0 \to \Lambda_c^{+} \overline{p}$~\cite{belle-lambda-pbar}, we search for charmless two-body baryonic decays, $B^0 \to p \overline{p}$, $B^0 \to \Lambda \overline{\Lambda}$, and $B^+\to p \overline{\Lambda}$. 
Previous work on these channels set upper limits on the branching fractions at the ${\cal O}(10^{-6})$ level~\cite{experiment}. 

Unexpected, large, transverse polarization has recently been
measured in $B \to \phi K^*$ decay~\cite{phiK*}. Measuring the
polarization in two-body baryonic decays has been
proposed~\cite{theory-part2} as a means to shed light on its
origin. This could provide the key to understanding whether there is
new physics in charmless $B\to VV$ decays.
  
Recent observations of three-body $B$ decays containing baryons in the final states~\cite{belle-threebody} suggest that production is enhanced at low invariant masses of the baryon system. Some theorists suggest that baryon production is favored by the reduced energy release on the baryon side~\cite{HouSoni}. The rates for three-body baryonic decays are large because the baryonic daughters can form more readily when an energetic meson is recoiling against the di-baryon system. In this paper we study the complementary suppression of two-body baryonic decays. The results presented here are based on a $140\,{\rm fb}^{-1}$ data set, corresponding to 152 million $B\overline{B}$ pairs, collected with the Belle detector at KEKB~\cite{KEKB}, an asymmetric $e^+e^-$ collider operating at the
$\Upsilon$(4S) resonance~\cite{charge-conjugate}.

 
The Belle detector is a large-solid-angle magnetic spectrometer that
consists of a three-layer silicon vertex detector (SVD), a 50-layer
central drift chamber (CDC), an array of aerogel threshold Cherenkov
counters (ACC), time-of-flight scintillation counters (TOF), and an
array of CsI(Tl) crystals, all located inside a superconducting solenoid coil that provides a 1.5 T
magnetic field.  An iron flux-return located outside of the coil is instrumented to detect $K_L^0$ mesons and to identify muons (KLM). The detector is described in detail elsewhere~\cite{Belle}.  
      
 
Tracking information is collected by the SVD and the CDC.
For each primary charged track, the impact parameter relative to the run-by-run interaction point (IP) is required to be within 2~cm along the $z$ axis (aligned opposite the positron beam) and within 0.05~cm in the plane transverse to this axis.

Particle identification (PID) for protons, kaons and pions is determined from the CDC specific ionization ($dE/dx$), the pulse-height information from the ACC and the timing information from the TOF. 
Proton candidates are selected based on $p/K/\pi$ likelihood functions obtained from the PID system. The selection criteria are $L_p/(L_p + L_K)$$>$ 0.6 and $L_p/(L_p + L_{\pi})$$>$ 0.6, where $L_p$, $L_K$, and $L_\pi$ represent the proton, kaon, and pion likelihoods, respectively. The proton detection efficiency is 70\%, determined from $\Lambda \to p \pi^-$ decays in the same momentum range as $B^0 \to p \overline{p}$ and $B^+\to p \overline{\Lambda}$. The corresponding misidentification rates for charged kaons and pions are 11\% and 4\%, determined using kaons and pions from the decay chain $D^{*-} \to \overline{D}{}^0 \pi^- \to (K^+ \pi^-)\pi^-$ in the above momentum range.

Candidate $\Lambda$ baryons are reconstructed in the $p\pi^-$ decay
 channel and are selected using the following requirements:
 1) the distance between two 
$\Lambda$ daughter tracks must be less than 12.9~cm along the $z$ axis and greater than 0.008~cm in the plane transverse to this axis;
 2) the flight length of the 
$\Lambda$ candidates must be greater than 0.22~cm in the plane transverse to the $z$ axis;
 3) the angular difference between the $\Lambda$ momentum vector and the vector to the decay vertex from the IP must be less than 0.09~rad~\cite{experiment}.
After application of the above selection criteria, the reconstructed 
invariant mass of the $\Lambda$ candidate is required to be between
1.111 and 1.121 GeV/$c^2$ (3$\sigma$). 


Candidate $B$ mesons are reconstructed from the primary charged tracks in the $p \overline{p}$ mode; from one primary charged track and one selected $\Lambda$ candidate in the $p \Lambda$ mode; from two selected $\Lambda$ candidates in the $\Lambda \overline{\Lambda}$ mode.
In these three modes, the $B$ meson candidates are selected using two kinematic variables defined in the $\Upsilon(4S)$ center-of-mass (CM) frame: the beam-energy constrained mass, $M_{\rm bc}=\sqrt{E^2_{\rm beam}-p^{*2}_{B}}$, and the energy difference, $\Delta E= E^*_B-E_{\rm beam}$, where $p^*_B$ and $E^*_B$ are the momentum and energy  of the $B$ candidate and $E_{\rm beam}$ is the beam energy.
The signal region is defined as $5.27\, {\rm GeV}/c^2 < M_{\rm bc} < 5.29\, {\rm GeV}/c^2$ and $|\Delta E|<$ 0.05 GeV.  The sideband region, used for background determination, is defined as $5.2\, {\rm GeV}/c^2 < M_{\rm bc} < 5.26\, {\rm GeV}/c^2$ and $|\Delta E|<$ 0.2 GeV.     


The decays $B^0 \to K^+ \pi^-$ and $B^0 \to \pi^+ \pi^-$ do not
contribute measurably to the background. This is due to their small branching
fractions, the small proton fake rates, and
the shift of the $B$ candidate out of the $\Delta {\it E}$ signal region. Background from other $B$
decays, studied using Monte Carlo (MC) simulation~\cite{Belle-MC}, is found to be negligible. 
The main background comes from continuum events, especially  $e^+e^- \to q \overline{q}\; (q=u,d,s)$. 

We use the same flavor tagging algorithm used in time dependent $CP$ violation measurements to make continuum suppression more effective.
Charged leptons, pions, and kaons that are not associated with the reconstructed $p \overline{p}$, $\Lambda \overline{\Lambda}$, or $p \overline{\Lambda}$ decay are used to identify the flavor of the accompanying $B$ meson. The algorithm in Ref.~\cite{flavor} introduces that the parameter $r$ is an event-by-event dilution factor ranging from $r = 0$ for no flavor discrimination to $r = 1$ for unambiguous flavor assignment.  
The signal and background
samples are separated into a group of lower tagging quality,
$0<r<0.75$ ($r_1$ region) and one of higher tagging quality, $0.75<r<1$ ($r_2$ region); these are treated separately in the following continuum background rejection method.  
 
For continuum suppression we use $\cos {\theta_T}$, which is the cosine of
the angle between the direction of the primary proton (for $p \overline{p}$ and $p \overline{\Lambda}$ modes) and the thrust axis~\cite{thrust} of the non-candidate tracks and showers.  For the $\Lambda \overline{\Lambda}$ mode, it is the cosine of the angle between the direction of the $\Lambda$ candidate and the thrust axis of the non-candidate tracks and showers.
We preselect events by requiring $\cos {\theta_T}$ to be less than 0.9.

We use seven variables to characterize the event topology: five
modified Fox-Wolfram moments~\cite{sfw}, $S_\perp$~\cite{sperp} and 
$|\cos {\theta_T}|$. We define a Fisher discriminant $\mathcal {F}$~\cite{fisher}, which is the sum of these seven variables with coefficients optimized to separate signal and background events. Probability density functions (PDFs) for the signal and for background as functions of $\mathcal {F}$ and of the cosine of the polar angle ($\theta_B$) of the $B$ candidate's flight direction are obtained.
An additional variable, $\Delta z$, is used: the decay
vertex fits are performed using the charged tracks from each of the $B$ decays. The distance between the two $B$ decay vertices in the $z$
direction, $\Delta z$, is obtained. For $B \overline{B}$ events the $\Delta
z$ distribution is wider than for random combinations of proton jets
from the continuum, which are studied in the MC and sideband data.  We
use a double Gaussian function for  $\Delta z$ as the fitted PDF if
$|\Delta z|<$ 0.2~cm. For events outside this $\Delta z$
range, we do not use the $\Delta z$ PDF for further calculations.

The signal and background likelihoods, $\mathcal {L}_S$ and $\mathcal
{L}_B$, are formed from the product of the PDFs for $\mathcal {F}$ and
$\cos \theta_B$ and, for the $p\overline{p}$ mode only, $\Delta z$. We
demand that $\mathcal {R}$ = $\mathcal {L}_S/(\mathcal {L}_S +
\mathcal {L}_B)$ exceeds 0.8 (0.85 for the $p\overline{p}$ mode) in the $r_1$
region and 0.2 in the $r_2$ region. This criterion is chosen to
optimize the figure-of-merit ($N_S/\sqrt{N_B}$) calculated using both MC
and sideband data samples, where $N_S$ and $N_B$ are predicted signal
and background yields, respectively. We optimize the $\cal {R}$
requirement by assuming that the branching fractions
for $B^0 \to p \overline{p}$, $\Lambda \overline{\Lambda}$, and $B^+\to p
\overline{\Lambda}$ are 1.0 $\times$ $10^{-7}$. We obtain the expected signal
yields from the product of the total number of $B\overline{B}$ events, the
signal efficiency from MC and the assumed branching fractions. 
The predicted background yields are obtained by fitting the $\Delta {\it E}$ distribution for the sideband data.

\begin{figure}[!htb]
\begin{center}
\resizebox*{3.41in}{2.5in}{\includegraphics{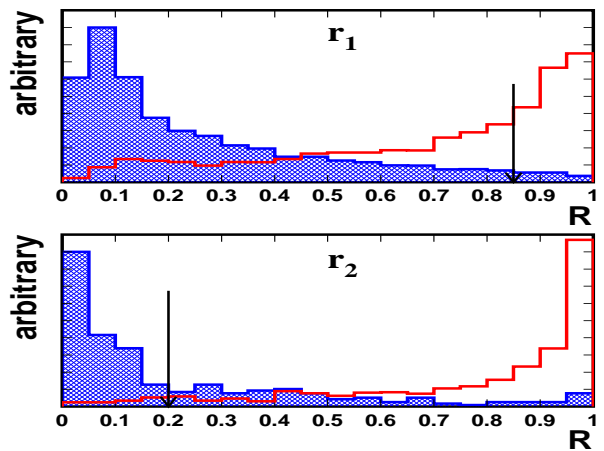}}
\end{center}
\caption{Distributions of $\mathcal {R}$ for events in the $r_1$ and
  $r_2$ regions. The open histogram is  $\mathcal {L}_S$ (signal MC) and the shaded histogram is $\mathcal {L}_B$ (sideband data). The arrows indicate the requirement positions.}
\label{fig:lr}
\end{figure}

We compare the sensitivities with and without the inclusion of the $\Delta z$ PDF in the $p \overline{p}$ mode. With the $\Delta z$ PDF included, the figure-of-merit improves by 26\% in the $r_1$ region and by 28\% in the $r_2$ region.
The $\mathcal {R}$ distributions for events in the two regions of $r$ are shown in Fig.~\ref{fig:lr}. 

The signal efficiency for each mode, after application of all the
selection criteria, is determined by MC simulation and itemized in
Table~\ref{tab:eff}. The systematic error in this efficiency arises
from tracking efficiency (2.0\%--4.7\%), proton identification (0.7\%--0.8\%
per proton, measured from a study of $\Lambda$ decays), $\Lambda$
selection (2.5\% per $\Lambda$, including all of the optimized
$\Lambda$ selection requirements from a study of $\Lambda$ decays), and the
likelihood ratio requirement in the two flavor-tagged regions (3.0\%--4.6\%
in the $r_1$ region and 4.4\%--7.7\% in the $r_2$ region, based on a study
of $B^- \to D^0 \pi^- \to (K^- \pi^+) \pi^-$ and $\overline{B}{}^0 \to D^+ \pi^-
\to (K^- \pi^+ \pi^+) \pi^-$). By comparing MC and data, corrections are made for the effect of
proton identification ($-4.0\%$, from a study of $\Lambda$ decays), the likelihood ratio requirement ($-4.5$\% to 0.8\%
in the $r_1$ region and 1.5\% to 3.2\% in the $r_2$ region, based on a study of $B^- \to D^0 \pi^- \to (K^- \pi^+) \pi^-$ and $B^0 \to D^+ \pi^-
\to (K^- \pi^+ \pi^+) \pi^-$), and polarization ($-2.2\%$ for the $B^0 \to \Lambda \overline{\Lambda}$ mode, due to the correlation of the spins of ($p, \Lambda$) and ($\overline{p}, \overline{\Lambda}$)).

 The total systematic uncertainties are 5.1\%, 9.2\% and 5.7\% for the $p \overline{p}$, $\Lambda \overline{\Lambda}$ and $p \overline{\Lambda}$ mode, respectively. 


By fitting
the $\Delta {\it E}$ sideband data (0.05 GeV $<$ $|\Delta \rm E|$ $<$ 0.2
GeV), projected on the $M_{\rm bc}$ axis ($5.2\, {\rm GeV}/c^2 < M_{\rm
  bc} < 5.3\, {\rm GeV}/c^2$), we obtain the background shape of $M_{\rm
  bc}$. By modeling the $M_{\rm bc}$ distribution with an ARGUS~\cite{Argus}
function and assuming a linear shape for $\Delta {\it E}$, we obtain the
predicted background by scaling the number of data events in the
sideband region to the signal region. 

\begin{figure}[!htb]
\begin{center}
\resizebox*{3.5in}{2.8in}{\includegraphics{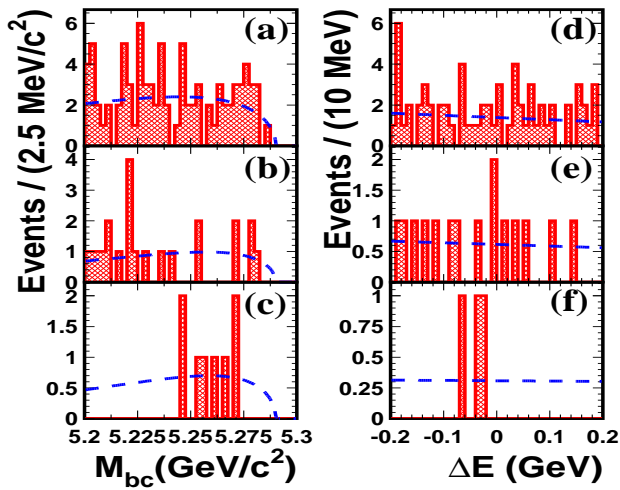}}
\end{center}
\caption{Distributions of $M_{\rm bc}$ and $\Delta {\it E}$ for (a)(d)
  $B^0 \to p\overline{p}$, (b)(e) $B^+ \to p \overline{\Lambda}$ and (c)(f) $B^0
  \to \Lambda \overline{\Lambda}$ candidates. We require the $\Delta {\it E}$ signal selection to plot the $M_{\rm bc}$ distribution and vice versa. The dashed lines represent the fitted background curves.}
\label{fig:data}
\end{figure}

Since few signal candidates are found (Fig.~\ref{fig:data}), we
determine the 90\% confidence level upper limits for the branching fractions by an extension of the Feldman-Cousins method~\cite{pole}.  The final results are listed in Table~\ref{tab:eff}.

\begin{table}[]
\begin{footnotesize}
\caption{Summary of the $B^0 \to p \overline{p},\, \Lambda \overline{\Lambda}$ and $B^+ \to p \overline{\Lambda}$ search, where $\epsilon$ is the reconstruction efficiency, $N_{obs}$ is the observed number of events in the signal region, $N^{bg}_{exp}$ is the expected background in the signal region and BF is the 90\% confidence level upper limit for the branching fractions. }
\label{tab:eff}
\begin{center}
\begin{tabular}{lrrrr}
\hline
\hline
Mode & $\epsilon$ [\%] & $N_{obs}$ & $N^{bg}_{exp}$ & BF [$10^{-7}$]\\
\hline
$B^0 \to p \overline{p}$ & 20.28 $\pm$ 1.03 & 17 & 14.9$\pm$3.4 & $< 4.1$ \\
$B^0 \to \Lambda \overline{\Lambda}$ & 4.32 $\pm$ 0.40 & 2 & 1.2$\pm$0.3 & $< 6.9$ \\
$B^+ \to p \overline{\Lambda}$ & 9.22 $\pm$ 0.53 & 5 & 3.5$\pm$1.1 & $< 4.9$ \\
\hline \hline
\end{tabular}
\end{center}
\end{footnotesize}
\end{table}                         

The results reported here use 4.8 times more $B\overline{B}$ pairs than the previous analysis~\cite{experiment}. 
Upper limits on the branching fractions at 90\% confidence level are   4.1 $\times$ $10^{-7}$,  6.9 $\times$ $10^{-7}$, and  4.9 $\times$ $10^{-7}$ for the $p \overline{p}$, $\Lambda \overline{\Lambda}$ and $p \overline{\Lambda}$ mode, respectively. 
The upper limits are reduced by factors of 2.9, 1.9 and 4.5. 
These results are consistent with constraints obtained by the other collaborations~\cite{current-result}, and indicate that $B$ decays to charmless di-baryon systems are suppressed.

In conclusion, we find no significant signal for the modes $B^0 \to p \overline{p}$, $\Lambda \overline{\Lambda}$ and $B^+\to p \overline{\Lambda}$ in 152 million  $B \overline{B}$ events.

We thank the KEKB group for the excellent operation of the
accelerator, the KEK cryogenics group for the efficient
operation of the solenoid, and the KEK computer group and
the NII for valuable computing and Super-SINET network
support.  We acknowledge support from MEXT and JSPS (Japan);
ARC and DEST (Australia); NSFC (contract No.~10175071,
China); DST (India); the BK21 program of MOEHRD and the CHEP
SRC program of KOSEF (Korea); KBN (contract No.~2P03B 01324,
Poland); MIST (Russia); MHEST (Slovenia);  SNSF (Switzerland); NSC and MOE
(Taiwan); and DOE (USA).



\begin{thebibliography}{99}
\bibitem{charge-conjugate} Charge conjugate modes are implicitly included throughout this paper.
\end{thebibliography}
\end{document}